\documentclass{aa}
\usepackage{graphicx}

\input{epsf}

\def\be{\begin{equation}}
\def\ee{\end{equation}}
\def\ba{\begin{eqnarray}}
\def\ea{\end{eqnarray}}

\def\msun{M_\odot}

\def\ltsima{$\; \buildrel < \over \sim \;$}
\def\simlt{\lower.5ex\hbox{\ltsima}}
\def\gtsima{$\; \buildrel > \over \sim \;$}
\def\simgt{\lower.5ex\hbox{\gtsima}}

\begin{document}

\title{ HI holes in galactic disks: Tracing the dark matter distribution}
\titlerunning{HI holes in galactic disks}
\author{E. I. Vorobyov\inst{1} 
and Yu. A. Shchekinov\inst{2}\\ 
}

\offprints{E. I. Vorobyov}
\institute{ Institute of Physics, Stachki 194, Rostov-on-Don, Russia and Isaac
Newton Institute of Chile, Rostov-on-Don Branch \\
\email{eduard\_vorobev@mail.ru}
\and
Department of Physics, University of Rostov, 
and Isaac Newton Institute of Chile, Rostov-on-Don Branch,\\
Rostov-on-Don, 344090 Russia\\
\email{yus@phys.rsu.ru}
}
\date{}

\abstract{ Multiple SN explosions in disk galaxies efficiently evacuate gas and 
form cavities with the sizes and shapes of the surrounding envelopes determined by the 
total amount of injected energy and by the initial gas distribution. Such 
cavities are seen as HI holes when observed in face-on galaxies.  
Gas hydrodynamics simulations are performed to obtain the quantitative 
characteristics of HI holes that could serve for the determination of the gas 
vertical scale height and the corresponding dark matter content and its 
distribution. Among these characteristics is the ratio of the maximum column 
density in the HI ring surrounding the hole to the background HI column density 
and the vertical expansion velocity of gas in the HI ring. We show that in some 
cases the extragalactic background ionizing radiation may produce HI holes in 
the outer regions of galaxies, and can account for the existence of HI holes in 
nearby face-on galaxies with the apparent lack of an underlying stellar 
population. 
}
\maketitle

\section{ Introduction}
The interstellar medium (ISM) shaped by the energy release from exploding 
massive stars provides the best sample for studying star formation processes on 
a ``microscopic'' level. The diversity of HI structures and associated H$\alpha$ 
distributions discovered in our own and nearby galaxies shows that star 
formation processes and dynamics of interaction of massive stars with the ISM 
are strongly environmentally dependent. Prominent examples are found in the 
Milky Way (Heiles 1984), 
M31 (Brinks \cite{Brinks}, Brinks \& Bajaja \cite{BB}), M33 (Deul \& den 
Hartog \cite{Deul}), Holmberg II (Puche et al. \cite{Puche}), NGC 891 
(Dettmar \& Schulz \cite{Dettmar1}), M101 and NGC 6946 (Kamphuis \cite{Kam}), 
NGC 253 (Sofue, Wakamatsu \& Malin \cite{Sofue}), 
SMC (Stavely-Smith et al. \cite{SS}) and LMC (Kim et al. \cite{Kim}), 
IC 10 (Wilcots \& Miller \cite{Wilcots}), IC 2574 (Walter \& Brinks \cite{Walter}), 
Holmberg I (Ott et al. \cite{Ott}), and others. 
Edge-on (such as NGC 891, NGC 4631, NGC 253) and face-on (such as Holmberg I 
and II, IC 2574, SMC and LMC) galaxies show these 
``microscopic'' structures in vertical and horizontal directions, respectively, 
and develop our understanding of not only the 
interrelation between the observed HI/H$\alpha$ structures and star formation 
activity (Dettmar \cite{Dettmar2}, Rand \cite{Rand}, Sofue, Wakamatsu \& Malin 
\cite{Sofue}, Rossa \& Dettmar 2003, Stewart et 
al. \cite{Stewart}, Walter \& Brinks \cite{Walter}), but also the factors governing the star formation process itself. 

Among these factors, the dark matter distribution in radial and vertical directions is of 
principle importance. 
At the same time, the influence of the dark matter distribution on the 
dynamics of HI and H$\alpha$ structures in edge-on and face-on galaxies was 
left aside general discussion.
Only recently Ott et al. (2001) paid closer attention to the relevance 
of HI dynamics in the dwarf irregular galaxy Holmberg I to the vertical gas 
distribution. However, available estimates of the vertical scale heights are 
still based on qualitative arguments and assumptions on the contribution of the dark matter to the total gravitational potential. 
In this paper we address the question of whether the observations of morphologies and quantitative characteristics of the HI holes in face-on galaxies can 
provide us with firm conclusions about the vertical scale height of the HI distribution. 
For this purpose, we study numerically the characteristics of HI flows associated with 
multiple SN explosions. 
We determine the dependence of these characteristics on the vertical scale height of 
the HI distribution and, consequently, on the vertical shape of the gravitational potential and dark matter distribution. 

The paper is organized as follows. In Sect.~\ref{model} the numerical model
is formulated, and the quantitative characteristics of HI holes are numerically
investigated for different vertical scale heights of gas distribution in
face-on galaxies. In Sect.~\ref{disc} the influence of the extragalactic
background ionizing radiation on the vertical shape of the HI distribution 
is discussed. The main results are summarized in Sect.~\ref{sum}. 

\section{SNe driven supershells: dynamics and morphology}
\label{model}
\subsection{Model }
  
The equilibrium HI distribution is obtained by solving the steady-state momentum equation  in cylindrical $(r,z)$ coordinates in a fixed gravitational 
potential determined by the stellar  
and dark matter components. The self-gravity of gas is neglected. 
The dark matter density distribution ($\rho_{\rm d}$)
is  assumed to be that of a modified isothermal sphere (Binney \& Tremaine
\cite{Binney}) 
\be
\rho_{\rm d}={\rho_{\rm d0}\over (1+r/r_0)^2}, 
\ee
where $r$ is the radial distance from the galactic center, 
$\rho_{\rm d0}$ is the central density, and $r_0$ is the characteristic 
scale length.  
The stellar component is assumed to be vertically stratified (Spitzer \cite{Spitzer})
\be
\rho_\ast=\rho_{\ast 0}\:{\rm sech}^{2}(z/h_\ast),
\ee
where $\rho_\ast$ is the stellar density, $\rho_{\ast 0}$ is the stellar midplane
density, and $h_\ast$ is the vertical scale height of stellar distribution.

\begin{table}
      \caption[]{Model parameters}
      $$
      \begin{array}{p{0.1\linewidth}l}
      \hline
      \noalign{\smallskip}
      $h^\dagger$ &\quad \rho_{d0} ~\quad 
~r_0~\quad ~~\rho_\ast~\quad ~~h_\ast~ 
     \quad ~~v_\ast ~~\quad ~M_d\\
      \hline
      \noalign{\smallskip}
      0.5 &\quad 0.07~\quad 0.227\quad 0.02~~\quad 0.3~\quad 7~\quad~~8\times 10^7\\
      \noalign{\smallskip}
      0.37 &\quad 0.015\quad 2.26~\quad 0.02~~\quad 0.3~\quad 7~\quad ~~3.7\times 10^8\\
      \noalign{\smallskip}
      0.21 &\quad 0.13~~\quad 1.5~~\quad 0.03~~\quad 0.3~\quad 8.5\quad~2.2\times 10^9\\
      \noalign{\smallskip}
      \hline
      \end{array}
      $$
\begin{list}{}{}
\item[$^{\mathrm{\dagger}}$] all scales are in kpc, densities in $\msun$ 
pc$^{-3}$, stellar  velocity dispersion ($v_{*}$)  in km s$^{-1}$, the
mass of the dark matter ($M_{\rm d}$) in $\msun$ is computed within the radius of 2 kpc.
\end{list}
\end{table}
The resulting equilibrium HI distribution is distinct for different dark matter and stellar 
distributions. Its vertical shape is neither Gaussian nor exponential (Celnik et al. \cite{Celnik}), however, 
it is better fitted by a Gaussian than by an exponent.
We consider three different models with three different gas vertical scale heights. 
The parameters of the stellar and dark matter distributions for each model
are shown in Table~1. Fitting the equilibrium vertical HI distribution
of each model by a Gaussian, we obtain the 1~$\sigma$ gas 
scale heights of $h$=210, 370, and 500~pc at the galactocentric radius of 2~kpc. 
For comparison, the corresponding exponential gas scale heights are $h_{\rm
exp}$=230, 400, and 540~pc, respectively.

Further, we assume that the size of SN-driven shells is much smaller than that of a galaxy. The radial dependence of the dark matter gravitational potential within the shell can therefore be neglected and the initial equilibrium
configuration of gas becomes a function of only the distance above the midplane
of a galaxy. In all models the initial gas velocity dispersion was taken 
as $v_{\rm g}=9$ km s$^{-1}$. 

The energy of supernova explosions is released in a sphere
with a radius of four zones. We use the constant wind 
approximation described in detail in Mac Low \& Ferrara (\cite{MF}).
We convert the energy of each SN explosion ($10^{51}$~ergs) totally into 
the thermal energy, because in the present simulations we deal with large stellar 
clusters capable of producing a hundred supernovae. 
With such a number of SN explosions, 
the surrounding ISM  will be quickly heated and diluted, making radiative 
cooling within the injection sphere ineffective. We choose the energy input phase to last for 30~Myr, 
which roughly corresponds to a difference in the lifetimes of the most and 
least massive stars capable of producing SNe in a cluster of simultaneously 
born stars.

SNe generate a supersonically expanding wind that compresses 
the gas, thus creating a bubble filled with the hot ejected gas 
surrounded by a shell of compressed cold material.
The gas dynamics is followed by solving the usual set of 
hydrodynamical equations in cylindrical coordinates 
using the method of finite-differences with a time-explicit, operator split
solution procedure of the ZEUS-2D numerical hydrodynamics code
described in detail in Stone \& Norman (\cite{Stone}). We have implemented the cooling curve of B\"ohringer \& Hensler (\cite{BH}) for a metallicity of one tenth of solar.
We use an empirical heating function tuned to balance the cooling in 
the background atmosphere so that it maintains the gas in hydrostatic 
equilibrium and may be thought of as a crude model for the stellar energy 
input. 
Cooling and heating are treated numerically using Newton-Raphson 
iterations, supplemented by a bisection algorithm for occasional zones where 
the Newton-Raphson method does not converge. In order to monitor accuracy, the 
total change in the internal energy density in one time step is kept below 
$15\%$. If this condition is not met, the time step is reduced and a solution 
is again sought.

\subsection{Qualitative description}
\label{descr}

\begin{figure}
  \resizebox{\hsize}{!}{\includegraphics{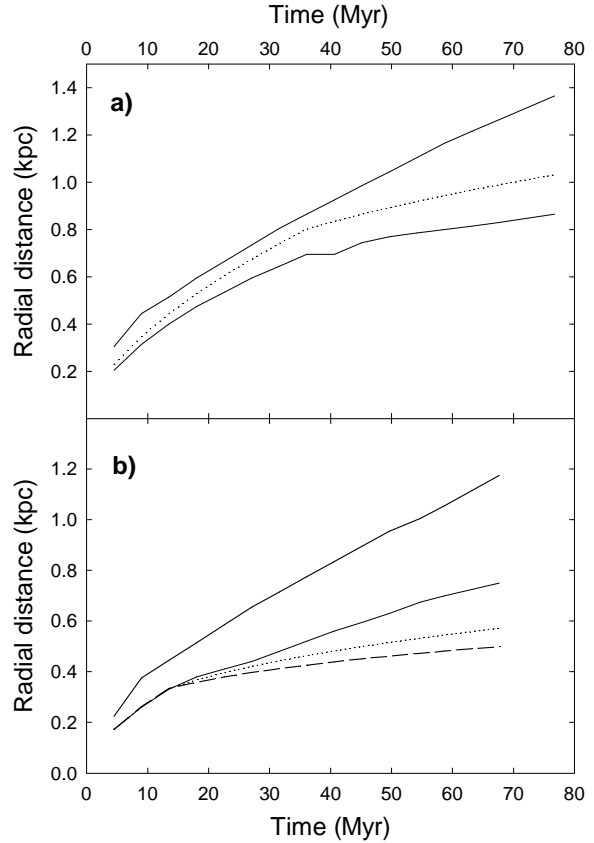}}
      \caption{{\bf a)} The solid lines delineate the inner and outer radii 
      of the shell expanding in a galaxy with $h=500$ pc. 
The dotted line gives the radius of the shell
as predicted from  the $R \propto (L/\rho)^{1/5}\: t^{3/5}$ law for the
SN-driven expansion at $t<35$~Myr (Castor et al. \cite{Castor}) and  
from the $R\propto t^{1/3}$ law for the "snow-plow" expansion at $t>35$~Myr. 
{\bf b)} The same for $h=210$ pc. The dotted line gives the radius 
of the shell
as predicted from  the $R\propto t^{1/3}$ law for a cylindrical ``snowplow'' 
expansion at $t>13.5$~Myr, the dashed line shows a spherical ``snowplow'' 
$R\propto t^{1/4}$ expansion. }
         \label{fig1}
\end{figure}

At the initial stages the remnant comes into a radiative phase 
relatively early, approximately at 
\be
t_r\sim 10^5\left({L_{38}\over n}\right)^{1/2}~{\rm yr},
\ee
when its radius is only 
\be
R\sim 25\left({L_{38}\over n}\right)^{1/2} ~{\rm pc},
\ee
where $L_{38}$ is the mechanical luminosity in $10^{38}$~ergs~s$^{-1}$  
released by multiple SN explosions, $n$ is the ambient volume density. This implies 
that the remnant is always radiative before it expands out to $R\sim 2
h$ and
enters a breakthrough phase (i.e. the shell breaking out of the disk
and losing metal-enriched material to the intergalactic space). 
The breakthrough occurs at approximately 
\be
t\sim 10^{-3}\left({n\over L_{38}}\right)^{3/2}h^4~{\rm yr},
\ee
where $h$ is in parsecs. 
The further evolution shows a very fast drop of pressure inside the hot remnant
and the ``snowplow'' expansion of a quasi-cylindrical shell of compressed cold
material in the disk. Due to a sharp drop of pressure inside the remnant the 
expanding shell starts thickening and can no longer be treated as a thin shell.  
Figure 1a shows the temporal evolution of the inner and outer radii of the 
shell expanding in a face-on galaxy with a gas scale height of $h=500$ pc
and HI surface density of $\Sigma_{\rm HI} =10.7~M_{\odot}$~pc$^{-2}$.
The surface density of the swept-up gas is generally not constant 
across the shell, it reaches a maximum near the outer edge of the shell 
and falls off on both sides. We define the inner radius of the shell as 
a radial distance where $\Sigma_{\rm HI}$ drops to its unperturbed value 
in the inner side of the shell. 
The outer radius is defined in the same manner. The mean radius is 
then obtained as an arithmetic average of the inner and outer radii.
The mechanical luminosity of $L_{38}=1$ is assumed, which is equivalent
to 100 SN explosions during the first 30~Myr.
The cooling rate for $T\geq 10^4$ K is taken from B\"ohringer \& Hensler 
(\cite{BH}), for $T<10^4$~K the cooling function rapidly declines to zero. 
It is seen that the shell remains thinner than a quarter or less of its 
mean radius only in the SN-driven expansion phase during the 
first 30-35 Myr.  At this evolutionary stage, 
the mean radius of the shell is well described by the $R \propto
(L/\rho)^{1/5}\: t^{3/5}$ law (Castor et al. \cite{Castor}) plotted by
the dotted line in Fig.~\ref{fig1}a, where
$\rho=0.5\:\rho_0$ and $\rho_{0}$ is the midplane mass density of the ambient
gas.
The shell grows very fast after $t=35$ Myr when a breakthrough is to occur. 
At the time of a breakthrough, the mean radius of the shell is $R \approx 1.6\:h$.
After the breakthrough the shell enters the ``snowplow'' phase and its mean radius grows approximately as $R\propto t^{1/3}$ 
(as shown in Fig.~\ref{fig1}a by the dotted line).

In case of a smaller gas scale height ($h=210$ pc), the breakthrough
occurs at t=13.5~Myr when $R \approx 1.8\: h$. The mean radius of the shell 
is poorly described by both the spherical ($R\propto t^{1/4}$, McCray \& Kafatos \cite{McCray}) and
cylindrical ($R\propto t^{1/3}$) thin-shell ``snowplow'' expansion laws
as shown in Fig.~\ref{fig1}b by the dashed and dotted lines, respectively.
Thus, it is obvious that when 
an HI hole seen in face-on galaxies is 
fitted by the spherical thin-shell ``snowplow'' expansion law $R\sim t^{1/4}$, the resultant 
energy input into the hole from SN explosions may be largely overestimated.   

\subsection{Dynamics versus vertical gas scale height}

At the early expansion stages, when the shells remain nearly 
spherical, their kinematical and dynamical characteristics 
scaled to the proper mechanical luminosity and local density are 
similar in different environments (see discussion in 
Castor et al. \cite{Castor}). Therefore, 
one should explore the later expansion stages in order to determine the large scale 
properties of a parent galaxy. At the later stages, the expanding 
shells are radiative and their expansion velocities are only weakly 
supersonic so that the corresponding Mach number is $M\simgt 1$ 
(in most face-on galaxies the expanding HI shells 
are marginally supersonic, or even subsonic as in the case of Holmberg~II, see 
discussion in Shchekinov et al. \cite{YUS}). In such conditions the compression 
factor of gas inside the shells (mass density contrast) is not high: for 
isothermal shock waves it is $\rho/\rho_0=\gamma M^2$, where $\gamma$ is 
the ratio of specific heats, and for weakly 
supersonic shocks at the later stages ($t\sim 30-40$ Myr) it can be around 
$\gamma$. 

A directly observed quantity is the HI column density. For an HI hole in 
a face-on galaxy created by consecutive SN explosions, 
the HI column density in a ring of compressed gas surrounding the hole is
\be
N({\rm HI})\simeq {2\sqrt{2}\over 3}{R^{3/2}n_0\over \sqrt{\Delta R}},
\label{number}
\ee 
where $R$ is the ring radius, $\Delta R$ is its thickness, $n_0$ is 
the midplane density of ambient gas $n_0=N_0({\rm HI})/h$, $N_0({\rm HI})$ is 
the column density of the background (unperturbed) HI gas. 
Substituting $\Delta R=R\:\rho_0/3\rho$ and $\rho/\rho_0\sim \gamma$ into
Eq.~(\ref{number}), one obtains a linear dependence of 
the relative HI column density on its radius in a face-on ring 
\be
{\cal K}_{\rm {\rm HI}}\equiv {N({\rm HI})\over N_0({\rm HI})}\simeq 2{R\over h},
\ee
with the slope inversely proportional to the HI vertical scale height of a parent 
galaxy. The highest relative column density ${\cal K}_{\rm HI}$ 
is reached when the breakthrough 
occurs at $R\sim \alpha\:h$, where $\alpha$ is a slowly varying function
of the vertical gas scale height. Simulations in Sect.~\ref{descr} indicate
that $\alpha \approx 1.6$ for $h=500$~pc, while for $h=210$~pc $\alpha \approx
1.8$. An expanding ring reaches maximal relative column density 
${\cal K}_{\rm HI}\sim 4$ only in galaxies with rather low vertical scale 
heights $h\le 200$~pc, where the compressed 
shell always remains supersonic. 
After the breakthrough, the compressed gas in the shell 
starts to move into the cavity because of a sharp drop in pressure, and 
the column density in the ring decreases approximately as $\propto R^{-1}$. 

When the gas scale height is sufficiently large, 
the shell expansion becomes sonic before breaking out of the disk, 
the associated perturbation propagates with the sound speed and simultaneously 
the shell itself starts swelling into the cavity, which 
results in a decrease of the gas column 
density. This occurs when
\be
R\simeq 230\:L_{38}^{1/2}\:h_{100}^{1/2}\:\Sigma_{10}^{-1/2}~{\rm pc},
\ee
where $h_{100}=h/100$ pc, $\Sigma_{10}$ is the unperturbed surface density
of gas ($\Sigma_{g0}$) in units of 10~$\msun$~pc$^{-2}$. 
For instance, when the gas scale height is $h=500$ pc and the unperturbed 
surface density $\Sigma_{g0}=10.5~\msun$ pc$^{-2}$, the shell becomes sonic 
when its radius is $R\simeq 500$ pc. As a consequence, 
the shell starts swelling into the
cavity and the corresponding relative column density of the ring decreases. 

\begin{figure}
  \resizebox{\hsize}{!}{\includegraphics{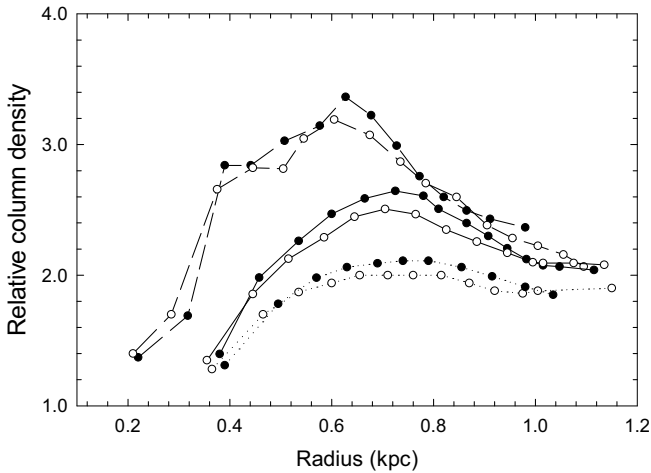}}
      \caption{Relative column density in the ring versus its radius: dashed,
      solid, and dotted lines represent $h$=~210,~500, and 700~pc, respectively; 
open circles correspond to $\Sigma_{g0}=15.2~\msun$ pc$^{-2}$, 
filled circles -- $\Sigma_{g0}=10.7~\msun$ pc$^{-2}$.}
         \label{fig2}
\end{figure}

Figure~\ref{fig2} shows the 
relative column density in the ring, ${\cal K}_{\rm HI}$, 
obtained in our numerical simulations for  three 
different vertical gas scale heights and two values of the 
undisturbed surface density $\Sigma_{g0}$. 
The curves exhibit the characteristic behaviour described above, with
the maximal values of ${\cal K}_{\rm HI}$ being higher for the lower 
vertical gas scale heights.
Specifically, for $h=210$~pc the maximum value of ${\cal K}_{\rm HI}$ is approximately
3.4 at the time when the shell starts swelling because of a sharp pressure
drop after the breakthrough. For larger $h$ the maximum value of ${\cal K}_{\rm
HI}$ is smaller. Note that ${\cal K}_{\rm HI}$ is virtually independent of
$\Sigma_{g0}$. 
For the total period of energy injection $\le 30$~Myr and 
for the vertical gas scale heights $<600$~pc, the maximum value 
of ${\cal K}_{\rm HI}$ and the corresponding radius of a shell
do not depend on the luminosity $L$, while in gaseous disks with $h\geq 600$ 
pc the maximum value of ${\cal K}_{\rm HI}$ and the corresponding radius scale as 
$L^{1/2}$. However, since a widely accepted value for the mechanical luminosity 
$L=10^{38}$~ergs~s$^{-1}$ is typical for OB~associations with a Salpeter IMF, 
one may expect that the curves depicted in Fig.~\ref{fig2} represent a
universal relation and may be used to draw firm conclusions about the vertical scale heights
of HI distributions in disk galaxies. 
It is obviously seen that the ${\cal K}_{\rm HI}$ versus $R$ relation is characteristic
for each vertical gas scale height only in the early phases  
when the shell has not yet broken out of the disk, while after the breakthrough 
${\cal K}_{\rm HI}$ can barely be distinguished among different $h$. 
Note in 
Fig.~\ref{fig2} a range in which the relative column density 
in the ring varies in galaxies with low vertical gas scale heights. 
It is 1.4-3.4 for $h=210$~pc, 
while  ${\cal K}_{\rm HI}$ is restricted to a narrower 
range of 1.3-2 in galaxies with $h=700$~pc. Thus, a narrow spread of observed 
column densities may indirectly indicate that the vertical gas scale height is large. 
Moreover, considering a universal and single-valued 
dependence of ${\cal K}_{\rm HI}(h,R)$ for $R\le 600-700$~pc, 
the local gas scale height $h$ can be 
inferred directly from the measurements of ${\cal K}_{\rm HI}$. 
Further, the measurements of ${\cal K}_{\rm HI}$ for the individual HI holes located at 
different galactocentric distances allow for the determination of the 
radial variations in the gas scale height. 

The increase in the mass of the dark matter generally makes a
gas disk thinner, i.e. the gas scale height $h$ decreases as
the dark matter mass increases.
Assuming the {\it local} plane-parallel gas distribution, the total density
$\rho_{\rm tot}(0,r)$
in the disk at the galactocentric radius $r$ and  $z=0$ can be derived
from  $h(r)=v_{\rm g}/\sqrt{4\pi G \rho_{\rm tot}(0,r)}$ (van der Kruit
\cite{Kruit}). Provided that
the radial distribution of the visible mass is known from
independent measurements, one can obtain the radial distribution of the
dark matter in the galactic plane. These are, of course, very approximate
estimates; nevertheless, a similar approach applied to the dwarf
irregular galaxy Holmberg~I by Ott et al. (\cite{Ott}) has yielded a
dark matter mass within the HI content of the galaxy, 
the value of which is in agreement with that obtained in
a more sophisticated numerical modeling by Vorobyov
et al. (\cite{Vor}).
Note, however, that the flaring and warping of the gas disk may further
complicate this analysis.

\begin{figure}
  \resizebox{\hsize}{!}{\includegraphics{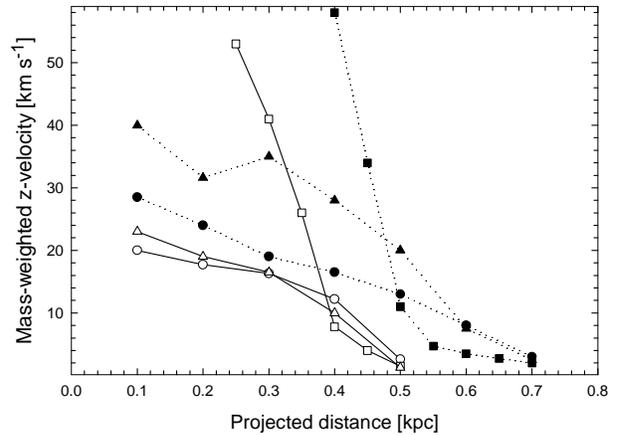}}
      \caption{Mass-weighted $z$-velocity profiles versus projected distance from
the geometrical center of the shells.
Solid and dotted lines represent the  shells with the radii of 0.5 and 0.7~kpc,
respectively. Open squares -- $h=210$~pc, $t=18$~Myr; open triangles --
$h=370$~pc, $t$=13.5~Myr; open circles -- $h=500$~pc, $t=13.5$~Myr.
Filled squares -- $h=210$~pc, $t=36$~Myr; filled triangles -- $h=370$~pc,
$t=31$~Myr; filled circles -- $h=500$~pc, $t=29$~Myr. 
The unperturbed HI surface density is $\Sigma_{g0}=10.7 \: M_{\odot}$~pc$^{-2}$. }
         \label{fig3}
\end{figure}

The velocity field of a shell expanding in a nearly face-on galaxy 
can additionally hint at the vertical structure of gas. 
At the initial stages far from breakthrough, the vertical component of gas velocity
in an expanding shell, $v_{\rm z}$, scales as $\sqrt{1-(r/R)^2}$, where $r$ is the 
projected distance from the geometrical center of a shell and $R$ is the shell radius. However, when 
the shell is near breakthrough, a substantial amount of gas constituting
the shell becomes involved in strong vertical motion. 
In Fig.~\ref{fig3} we plot
the mass-weighted $v_{\rm z}$ as a function of the projected distance from the 
geometrical center of a shell 
for different vertical gas scale heights and different fixed shell radii.
The solid lines represent the shells with a radius of $R=0.5$~kpc, while the
dotted lines correspond to the shells with a radius of $R=0.7$~kpc.

It is seen that the expanding gas is only weakly supersonic (the corresponding Mach number of 
2-3) in the shells that are at the early phases of expansion, well 
before breakthrough ($R=0.5$~kpc, $h=370$ and 500~pc). 
On the other hand, the shell that approaches the breakthrough phase 
($R=0.5$~kpc and $h=210$~pc) shows a steep velocity increase 
up to $v_{\rm z}=60$~km~s$^{-1}$
at the inner edge  of the projected shell. The $v_{\rm z}$ radial profiles
remain qualitatively similar at different phases of the shell expansion,
namely at $R=0.5$~kpc and $R=0.7$~kpc.

At the breakthrough phase, 
the radius of a shell is almost twice as large as 
the gas scale height $h$. 
From this point of view, one may expect that the most extended HI holes in 
face-on galaxies (such as HI holes number 2, 5, 8, 10 and others in Holmberg~II,
Puche et al. \cite{Puche} ) 
whose $z$-velocity reveals such a behaviour at the inner edges 
can provide us with a direct measure of the local gas scale height. 
Specifically, scanning the HI radial velocity 
profiles within a hole with a beam smaller than its size (for the
Ho~II case, the beam must be less than $4''-6''$) could help to
identify nearly breaking-through shells, the radii of which 
could then serve as a direct measure of the local gas scale height.

\begin{figure}
  \resizebox{\hsize}{!}{\includegraphics{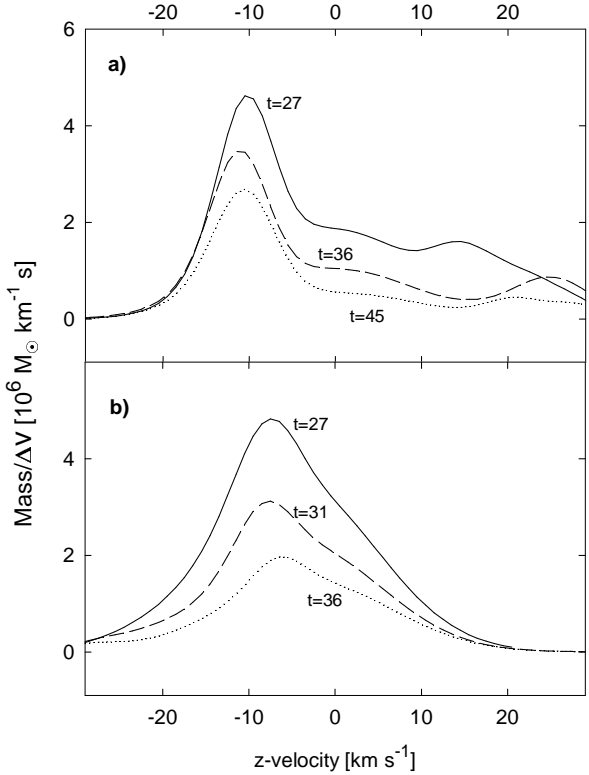}}
      \caption{The HI spectrum of an expanding shell 
produced by 100 successive SNe located 100~pc above the midplane
in a galaxy with {\bf a)} $h=500$~pc, {\bf b)} 
$h=210$~pc. The HI spectra are numerically obtained for 
a 1~kpc diameter beam centered on
the shell at three different times in Myr as indicated in each frame.}
\label{fig4}
\end{figure}

Out-of-plane SN explosions in nearly face-on galaxies can also provide quite a robust tool for the
determination of the gas vertical scale height. 
In Fig.~\ref{fig4} we plot typical HI spectra of the shells 
produced by 100 successive SNe located at 100~pc above the midplane. 
The HI spectrum is obtained
for a 1 kpc diameter beam centered on the expanding shell.
Two model galaxies with different gas scale heights are
considered: $h=500$~pc and $h=210$~pc. 
A one-dimensional velocity dispersion of $\sigma_{\rm g}=3$~km~s$^{-1}$ is assumed 
for the gas forming an expanding shell when constructing 
the model HI spectra, while  $\sigma_{\rm g}=9$~km~s$^{-1}$ is adopted for
the unperturbed gas. The HI spectrum of the shell expanding in a galaxy
with $h=500$~pc is characterized by a typical two-humped structure.
A smaller part of the accelerating shell expanding upwards 
($v_{\rm z}>0$) and a more massive part of the shell expanding downwards ($v_{\rm
z}<0$) with approximately 
constant velocity are clearly seen. The HI spectrum of the shell expanding in a galaxy
with $h=210$~pc has a single peak located at $v_{\rm z} \ne 0$.
Velocity centroids of gas expanding downwards are obviously 
smaller for lower gas scale heights $h$, because lower scale heights lead to a faster 
onset of breakthrough and, therefore, to a faster 
drop of pressure inside the SN-driven shell. 

\begin{figure}
  \resizebox{\hsize}{!}{\includegraphics{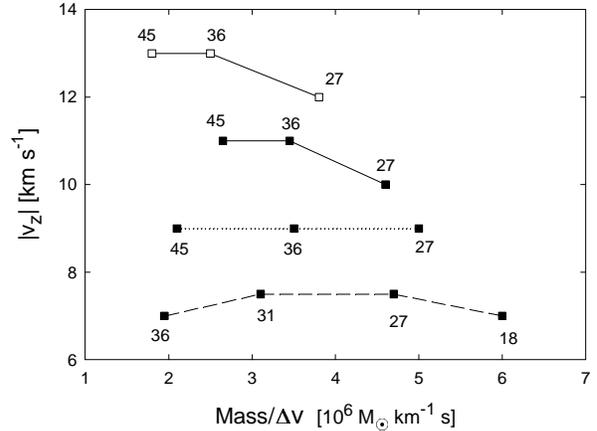}}
      \caption{The peak $z$-velocity versus HI intensity for different 
gas scale heights and locations of SN explosions above the midplane: 
dashed line -- $h=210$~pc, SNe at $z_\ast=100$~pc; dotted line
-- $h=370$~pc, SNe at $z_\ast=100$~pc; solid line with filled squares -- 
$h=500$~pc, SNe at $z_\ast=100$~pc; solid line with open squares -- 
$h=500$~pc, SNe at $z_\ast=50$~pc. 
Different evolutionary stages of the shells are labeled in Myr.}
\label{fig5}
\end{figure}

In Fig. \ref{fig5} we plot the absolute values of the peak $v_{\rm z}$ velocities of gas
expanding downwards as a function of the corresponding HI intensities (see
Fig.~\ref{fig4}) for three 
different gas scale heights: $h=210$, 370, and 500~pc. 
It is clearly 
seen that the smaller peak z-velocities are associated with the shells 
expanding in 
galaxies with lower gas scale heights. A distinguishing feature of this 
dependence is that the peak expansion velocity is virtually independent of 
intensity in a relatively wide range 
covering the latest stages of expansion before and after the breakthrough, 
and varies approximately as 
\be
v_z^{\rm peak}\simeq 2\left({h\over 100\:{\rm pc}}\right)^{0.9}\left({z_\ast\over 
100 \: {\rm pc}}\right)^{-0.24}~{\rm 
km~s^{-1}}.
\ee
where $z_{\ast}$ is the location of SNe above the midplane. 

\section{Effects of external ionizing radiation}
\label{disc}

One of the most striking facts related to HI holes in dwarf face-on galaxies
is the lack of an underlying stellar population 
that could produce a strong energy input into the ISM. This tendency 
is obviously seen 
in Holmberg II (Ho~II) where most extended HI holes are located outside the stellar 
disk (Rhode et al. \cite{Rhode}, Stewart et al. \cite{Stewart}). 
Hence, the SN origin of those HI holes in Ho~II is questionable.
An alternative explanation of the HI hole formation
connected with high-velocity HI clouds (HVCs) falling onto the disk seems to be 
excluded in the case of Ho~II, because no HVCs are found nearby. In this case, Efremov 
et al. (\cite{Efremov}) argue that a supernova in the merging event of a 
compact binary system may be responsible for HI hole creation. However, 
a distinguishing feature of multiple HI holes on the periphery of Ho~II is 
that they form quite a regular structure resembling a spiral wave (see 
Stewart et al. \cite{Stewart}), and seem unlikely to be produced by regularly placed 
supernovae explosions. In this connection it is worth mentioning the role of an
external ionizing radiation field in shaping the HI distribution, particularly on
the galactic periphery. 

\begin{figure}
  \resizebox{\hsize}{!}{\includegraphics{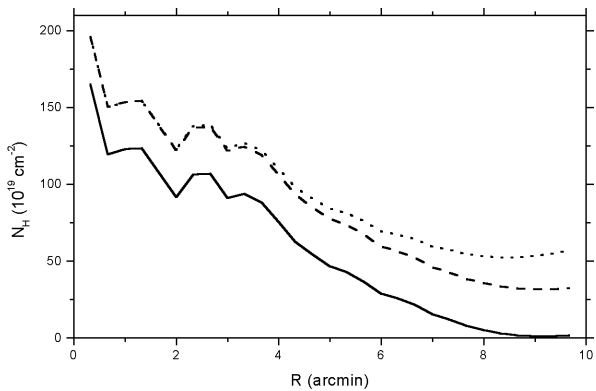}}
\caption{Comparison of the total hydrogen column density and the 
observed HI in Ho~II: solid line -- observed HI (Puche et. al. 1992), 
dashed line shows total column density $N_0({\rm H})$ from (\ref{tot}) with $h=625$ pc, 
dotted line -- $N_0({\rm H})$ with $h=625+18.6(R-1)^2$.  
}
\label{fig6}
\end{figure}

The position of the boundary between HI and HII 
layers, $z_i$, in an exponential vertical gas distribution 
ionized by a given flux of 
external UV photons $I$ is determined by the following equation 
$2\pi I=\alpha_rn_0^2
\int_{z_i}^\infty\exp(-2z/h)dz$. The observed column density of 
atomic hydrogen, $N({\rm HI})=2n_0\int_0^{z_i}\exp(-z/h)dz$, can then be found 
as 

\be
\label{tot}
N({\rm HI})=N_0({\rm H})-\sqrt{4\pi I h\over \alpha_r}, 
\ee
where $\alpha_r$ is the hydrogen recombination rate and $N_0({\rm H})=2n_0\:h$ 
is the total column density of hydrogen (both neutral and ionized), 
$n_0$ is the total midplane density of hydrogen. It is readily seen that 
for a typical value of the background ionizing flux of $I\sim 10^6$ photons 
cm$^{-2}$ s$^{-1}$ sr$^{-1}$ and for $h\sim 300$ pc the column density of ionized hydrogen can reach $3\times 
10^{20}$ cm$^{-2}$, nearly coincident with the observed in Ho II at the radii 
of $\sim 6$ arcmin (Puche et al. \cite{Puche}). This implies that the observed $N({\rm HI})$ 
may represent a small fraction of the total amount of hydrogen, 
and thus a relatively weak 
perturbation in $n_0$ and/or in $N_0({\rm H})$ can cause a rather strong variation in 
the amount of neutral hydrogen. This effect increases for a 
flaring gaseous disk with the scale height 
$h$ growing outwards. In Fig.~\ref{fig6} we plot the total column density 
of hydrogen, $N_0({\rm H})$, as a function of Ho~II radius. 
$N_0({\rm H})$ is derived from the observed column density
of neutral hydrogen $N({\rm HI})$ for two models of 
the vertical gas distribution in Ho~II: a plane-parallel distribution 
with $h$=625 pc and a flaring 
disk with $h=625+18.6(R-1)^2$ pc, where $R$ is given in arcminutes. 
In the outer regions at $R\simgt 6$ arcmin, the observed 
$N({\rm HI})$ may represent a relatively 
small (less than 50 \%) fraction of the total hydrogen content, 
and thus can be easily disturbed, ionized, and transformed into a hole 
of depleted HI column density. 

In principle, HI holes created by multiple SN explosions can 
become radially asymmetric due to radial variations in the gas density, 
particularly at 
larger distances from the galactic center where the flaring of the gas disk may dominate.
For an exponential disk with the density distribution 
$\rho\propto \exp(-z/h-R/R_0)$, the expansion velocity of a shell 
at a given point varies as $v_s\propto \exp(z/2h+R/2R_0)$. In this case, 
the geometrical center of an HI shell can be shifted outward with 
respect to the position of 
SN explosions  (Kalberla, private communication). Typical scales in 
vertical and radial directions are of the order of hundreds and thousands
of parsecs, respectively. For example, for the Ho~II galaxy $h\sim 600$ pc, while the 
radial scale for 
the HI surface density $R_0\sim 3$ kpc (Puche et al. 1992, Bureau \& 
Carignan, 2002). For face-on galaxies, direct measurements of the 
radial distributions of the  volume density 
or flaring are not available, but one can assume that the
radial scale lengths for them are of the same order as for the HI 
surface density. The corresponding deviations of the expansion velocities 
in radial $\delta v_R$ and vertical $\delta v_z$ directions from a precise 
spherical shape can be estimated then 
as $\delta v_R/\delta v_z\sim h/R_0\ll 1$. 
In the case of Ho~II, it becomes $\delta v_R/\delta v_z\sim 0.2$. 
Hence, one may conclude that the expected radial asymmetry of a shell, 
and the corresponding displacement of the geometrical center of a shell
with respect to the location of SN explosions, 
is less than $20\%$ of the gas scale height.

\section{ Summary}
\label{sum}

Clustered supernova explosions in the disk of a face-on galaxy produce
an expanding shell of compressed material, which is seen as an HI
ring surrounding a central HI depression.
In this paper we show that several physical characteristics of 
expanding shells are sensitive to the gas scale height and, hence, 
can be used for the
determination of the dark matter content in face-on disk galaxies. 

$\bullet$ A functional dependence of the relative column 
density of gas in the HI ring ($\cal K_{\rm HI}$)
on the radius of the ring is found to be specific for a given gas
scale height $h$, which makes it possible to infer $h$ from 
observations of even a single HI ring.
This is particularly true for the shells near
breakthrough phase. 

$\bullet$ The vertical component of gas velocity in an 
expanding shell reveals a characteristic behaviour (particularly, 
at the stages close to a breakthrough), which can be used to trace the 
gas scale height $h$ in galaxies with a sufficiently thin HI layer. 

$\bullet$ Out-of-plane SN explosions produce asymmetric expansion, with a
more massive part of the shell propagating towards the denser regions of the disk. 
The velocity of this part of the shell is a single-valued function of the scale 
height $h$ and the height above the midplane at which the SN explosions take place. 

$\bullet$ Some of HI holes, particularly in the outer regions of a galaxy, 
can be connected to the ionization of HI layer by extragalactic UV photons.

\section*{ Acknowledgments}
We would like to thank the referee, Dr. P. M. W. Kalberla, 
for his suggestions and critical comments that  substantially improved 
the paper.
This work was supported by the RFBR (projects No 00-02-17689) and
the INTAS grant YSF-2002-33. YS acknowledges financial 
      support from \emph{Deut\-sche For\-schungs\-ge\-mein\-schaft, DFG\/} 
      (project SFB N 591, TP A6).

\end{document}